\documentclass[a4paper,11pt]{article}
\usepackage{pos}
\usepackage{natbib}

\title{Ultra-high-energy cosmic ray acceleration by magnetic reconnection in relativistic jets and the origin of very high energy emission}
 \ShortTitle{UHECR acceleration by magnetic reconnection}

\author*[a]{Elisabete M. de Gouveia Dal Pino}
\author[b]{Tania E. Medina-Torrejon}
\author[a]{Luis Kadowaki}
\author[c]{Grzergoz Kowal}
\author[a]{Juan Carlos Rodriguez-Ramirez}
\affiliation[a]{Instituto de Astronomia, Geof\'{i}sica e Ci\^{e}ncias Atmosf\'{e}ricas, Departamento de Astronomia, Universidade de S\~{a}o Paulo,1226 Mat\~{a}o Street, S\~{a}o Paulo, CEP: 05508-090, Brazil}

\affiliation[b]{Instituto de F\'{i}sica, Universidade de S\~{a}o Paulo,1371 Mat\~{a}o Street, S\~{a}o Paulo, CEP: 05508-090, Brazil}

\affiliation[c]{Escola de Artes, Ci\^encias e Humanidades - Universidade de S\~ao Paulo,
Av. Arlindo B\'ettio, 1000 -- Vila Guaraciaba, CEP: 03828-000, São Paulo - SP, Brazil}

\emailAdd{dalpino@iag.usp.br}

\abstract{
Relativistic jets are believed to be born magnetically dominated. 
Very and ultra-high energy cosmic rays can be efficiently accelerated by magnetic reconnection in these sources. We here demonstrate this directly,  with no extrapolations to large scales, by means of three-dimensional relativistic magnetohydrodynamical (3D-RMHD) simulations of a Poyinting flux dominated  jet.  We inject thousands of  low-energy protons in the region of a relativistic jet that corresponds to the transition from magnetically to kinetically dominated, where its magnetization parameter is $\sigma \sim 1$. In this region, there is efficient  fast magnetic reconnection which is naturally driven by current-driven-kink instability (CDKI) turbulence  in the helical magnetic fields of the jet. We find that the particles are accelerated by Fermi process in the reconnection regions (and by drift in the final stages) up to energies $E \sim 10^{18}$  eV for background magnetic fields $B \sim 0.1$ G, and $E \sim 10^{20}$ eV for $B \sim 10$  G. We have also derived from the simulations the acceleration rate due to magnetic reconnection which has a weak dependence on the particles energy 
$ r_{acc} \propto  E^{-0.1}$, characteristic of exponential growth. The energy spectrum of the accelerated particles develops a power-law tail with spectral index $p \sim -1.2$. This hardness of the spectrum must decrease when particle losses and feedback into the background plasma are included. Our results can explain observed flux variability in the emission of blazars at the very high energy band  as well as the associated neutrino emission. Successful applications of our results  to the blazars MRK 421 and TXS 0506+056 are also discussed.
}

\FullConference{37$^{\rm{th}}$ International Cosmic Ray Conference (ICRC 2021)\\
		July 12th -- 23rd, 2021\\
		Online -- Berlin, Germany}

\begin{document}
\maketitle

\section{Introduction}

Relativistic Jets are born magnetically dominated 
and perhaps the most striking evidence on this has been recently provided by the EHT polarization measurements in the surrounds of the supermassive black hole of the active galaxy M87 \citep{akiyama_etal_2021a}.
Directly connected to this fact is that the
observations of several  blazars (active galactic nuclei, AGN, with highly beamed relativistic jets pointing to the line of sight) have evidenced very high energy (VHE) emission with time-variability of minutes in the TeV bands
(e.g., 
PKS 2155-304 \cite{aharonian_etal_07}),
implying extremely  compact and fast emission regions 
with Lorentz factors much larger than the typical  bulk values  in the jet of these sources ($\Gamma_{bulk} \simeq$ 5--10). 
This is very hard to be explained by standard particle acceleration in shocks, particularly if it comes from the magnetically dominated inner jet regions of these sources. Indeed, the only  mechanism  that seems to be able to explain both, the high variability and
compactness of the TeV emission in these sources is particle acceleration  in magnetic reconnection layers to PeV or larger energies 
\cite{giannios_etal_2009,medina_etal_2021}.  
Equally striking is the recent outstanding  simultaneous detection of gamma-rays and  high-energy neutrinos from the blazar TXS 0506$+$056
\cite{aartsen_etal_2018}. This has highlighted  the potential efficiency of AGN jets as ultra-high-energy cosmic ray (UHECR) accelerators, evidencing  for the first time the presence of high-energy protons interacting with ambient photons, producing  pions that subsequently  decay in gamma-rays and neutrinos.  It has been speculated  that if these protons are produced in the magnetically dominated regions of the jet near the source core, then they are most probably accelerated by fast magnetic reconnection \cite{dalpino_etal_2018}.

Particle acceleration by magnetic reconnection has been explored in a broad context from the solar system  \citep[e.g.][]{drake_etal_2006}  to jets and accretion disks in microquasars and AGNs
\citep[e.g.][and references therein]{
dalpino_lazarian_2005,giannios_etal_2009,
kadowaki_etal_15,singh_etal_15,
medina_etal_2021},  
pulsar wind nebulae 
\citep[e.g.][]{
cerutti_etal_2014},
and gamma-ray bursts (GRBs) \citep[e.g.][]{zhang_yan_11}.

Magnetic reconnection occurs when  magnetic fluxes of opposite polarity encounter, then partially break and rearrange their configuration at a velocity $V_{rec}$ which is  a substantial fraction of the local Alfv\'{e}n speed ($V_A$) if reconnection is 
fast \citep[e.g.][]{lazarian20}.
Different processes such as plasma instabilities, anomalous resistivity, and turbulence, may trigger fast reconnection. The latter process, in particular, is very efficient and probably the main driving mechanism of fast reconnection in large scale astrophysical flows \cite{lazarian_vishiniac_99}. 
Since turbulence causes an efficient field-fluid slippage and stochasticity of the magnetic field lines, bringing initially distant lines into close separations through Richardson diffusion 
\citep[][]{
lazarian20}, 
this leads to many patches reconnecting simultaneously, making the reconnection rate very fast (and actually independent of the intrinsic plasma microscopic magnetic resistivity and the Lundquist number). 

The rearrangement of the magnetic field configuration converts  magnetic into thermal and kinetic energies, 
and allows for efficient particle acceleration predominantly in a Fermi-like stochastic process \citep{dalpino_lazarian_2005} \citep[see also  reviews on this process, e.g. in][]{
dalpino_kowal_15}.

Particle acceleration in reconnection regions has been successfully tested in several numerical works employing multi-dimensional (2D and 3D)  magnetohydrodynamic (MHD) simulations with test particles \citep[e.g., ][]{kowal_etal_2011,kowal_etal_2012,delvalle_etal_16,
dalpino_etal_2018,
medina_etal_2021}, and particle-in-cell (PIC) simulations \citep[e.g.][and references therein]{drake_etal_2006,
sironi_spitkovsky_2014,
guo_etal_2016,
comisso_etal_2020}.

Both MHD and PIC simulations have probed the efficiency of particle acceleration by reconnection, but a fundamental difference is that while PIC simulations can probe only the microscopic scales of  the process up to  a few thousand of the plasma skin depth  ($c/\omega_{p}$) and early acceleration of the particles only up to a few $\sim 100$  times their rest mas ($m c^2$), the MHD simulations with test particles can probe the much larger scales of the astrophysical systems and accelerate the particles up to the observed UHEs, i.e.  several orders of the magnitude of their rest mass \cite{kowal_etal_2011,kowal_etal_2012,delvalle_etal_16,medina_etal_2021}.   

In this talk we summarize our recent results on the acceleration of  thousands of  test particles injected in a magnetized relativistic jet with the characteristics of a blazar jet, subject to current-driven kink instability (CDKI), based on  3D relativistic MHD simulations \cite{medina_etal_2021,kadowaki_etal_2021}. We find that the particles can be accelerated up to UHEs in the regions of fast magnetic reconnection which are developed in the turbulent flow  driven by the kink instability, thus demonstrating that these sources are really efficient UHECR accelerators \cite{medina_etal_2021}.  The magnetic reconnection power associated to the events of fast reconnection is able to produce a variability pattern that is consistent with observed flux variability at the gamma-ray band of blazars (e.g. MRK 421) \cite{kadowaki_etal_2021}. Finally, from these results we have also built a multi-zone lepton-hadronic model based on particle acceleration by magnetic reconnection  which is able to reproduce the spectral energy distribution (SED) of the  blazars, from BLLacs to FSRQs, including their VHE band and the neutrinos (Rodriguez-Ramirez et al., in prep.; see also Rodriguez-Ramirez et al. this Conference).

\section{Magnetic reconnection and particle acceleration in blazar jets}

Jets with strong helical or toroidal fields are subject to CDKI \citep[see e.g.][]
{singh_etal_16}. 
This instability distorts the jet column and induces non-helical currents that ultimately can break the magnetic field lines and drive magnetic reconnection. 

\begin{figure*} 
\centering
    \includegraphics[scale=0.285]{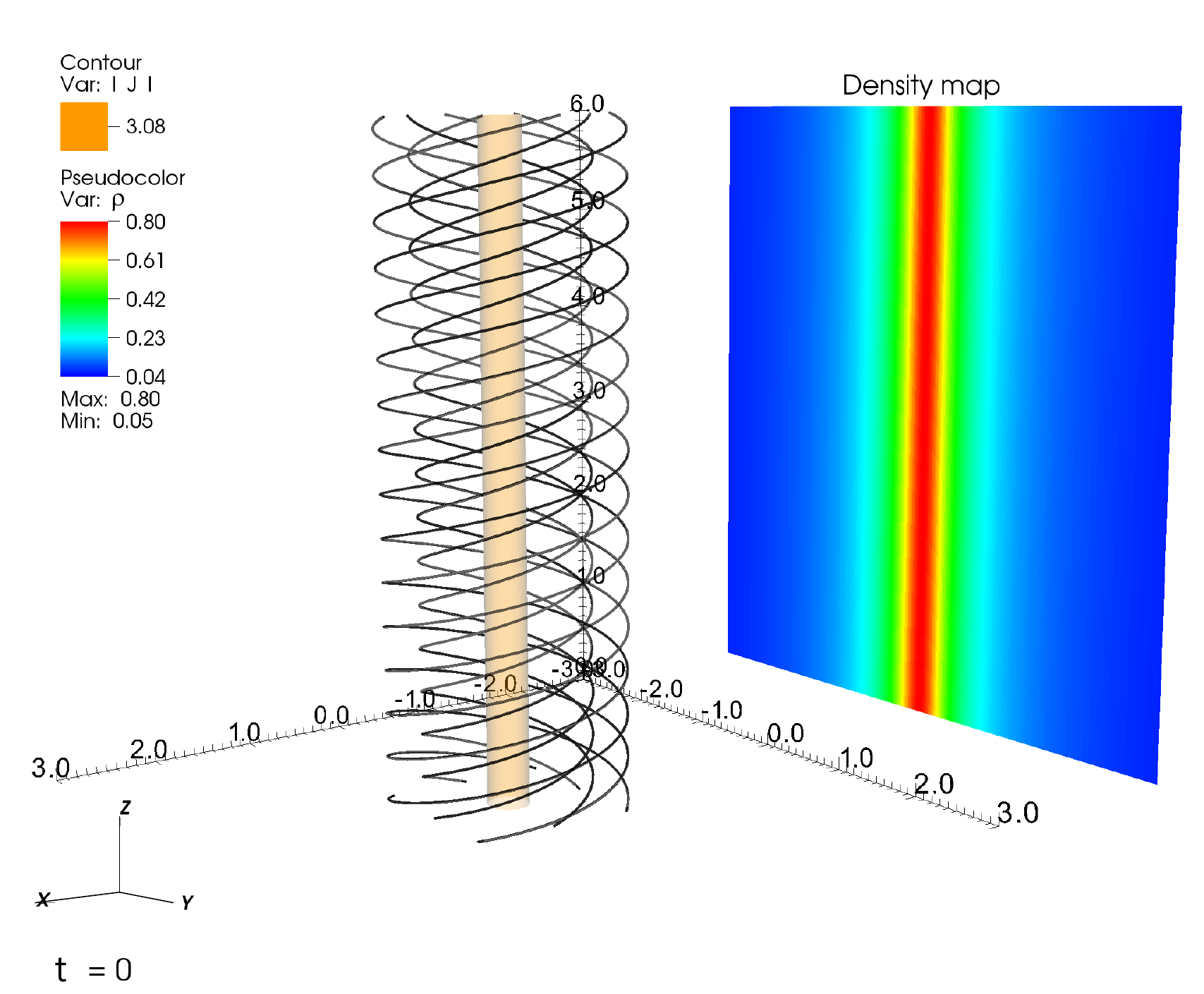}
	\includegraphics[scale=0.285]{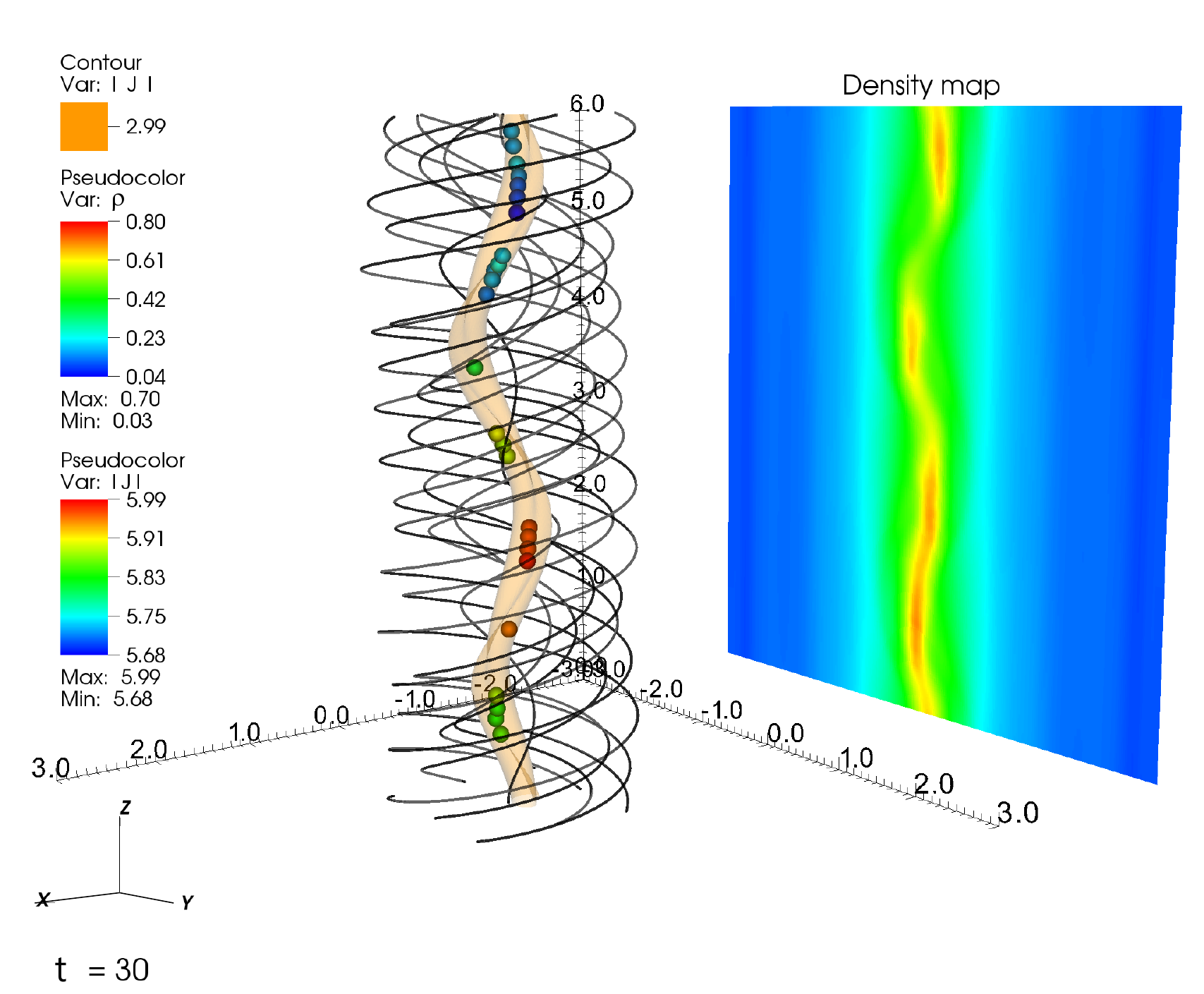}
	\includegraphics[scale=0.285]{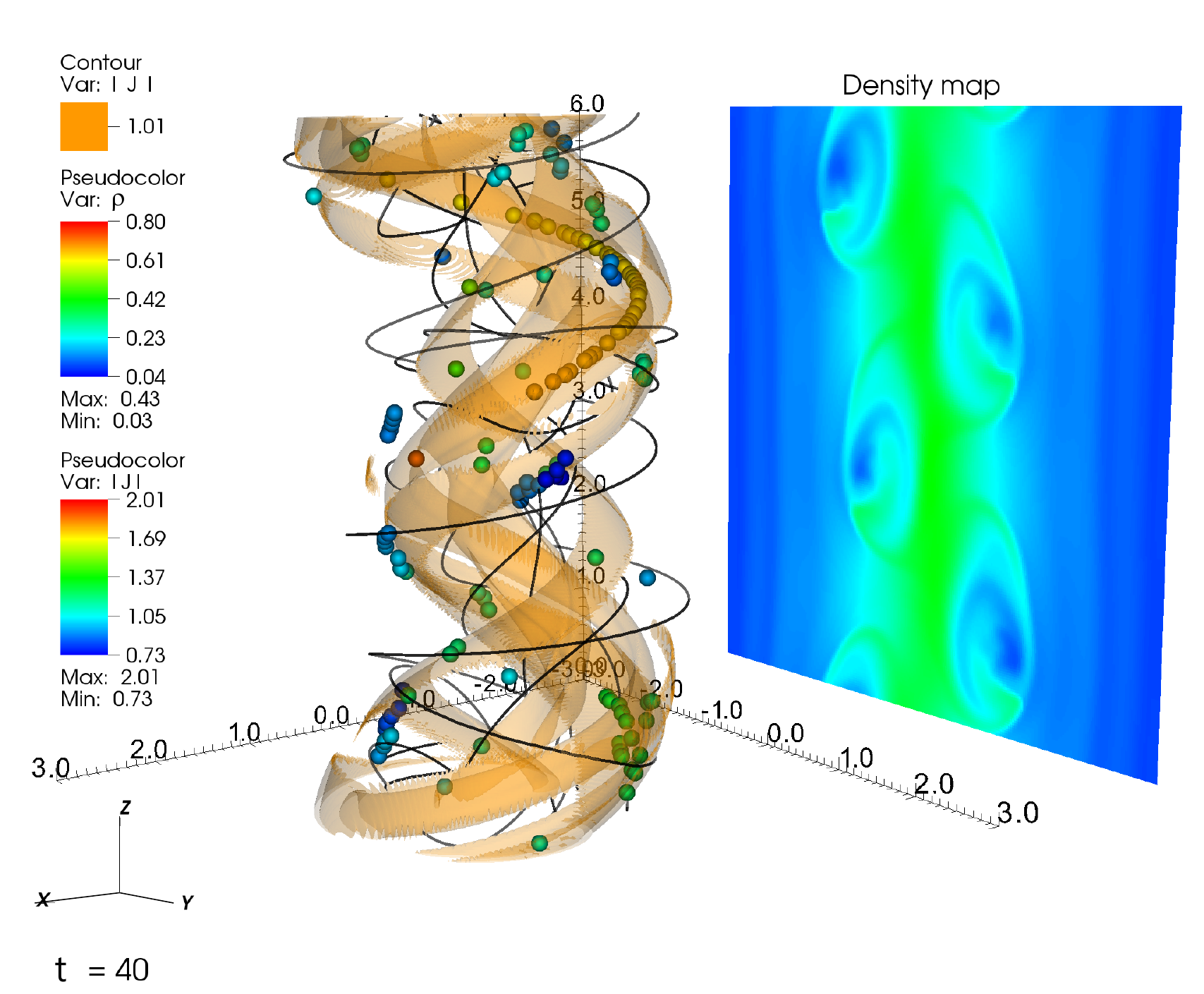}
\caption{
Time evolution of the jet at $t=0$, $30$ and $40$ $L/c$ (from left to right). The diagrams depict isosurfaces of the current density intensity at half maximum $|J|$ (orange color), the solid black lines correspond to the magnetic field lines, the density profile at the middle of the box ($y-z$ plane at $x=0$), and the magnetic reconnection sites identified by a search algorithm in the jet frame given by colored circles along the jet which  correspond to different current density magnitudes in code units. From \cite{medina_etal_2021,kadowaki_etal_2021}.}
\label{jet_points}
\end{figure*}
\begin{figure}
  \begin{center}
	\includegraphics[scale=0.45]{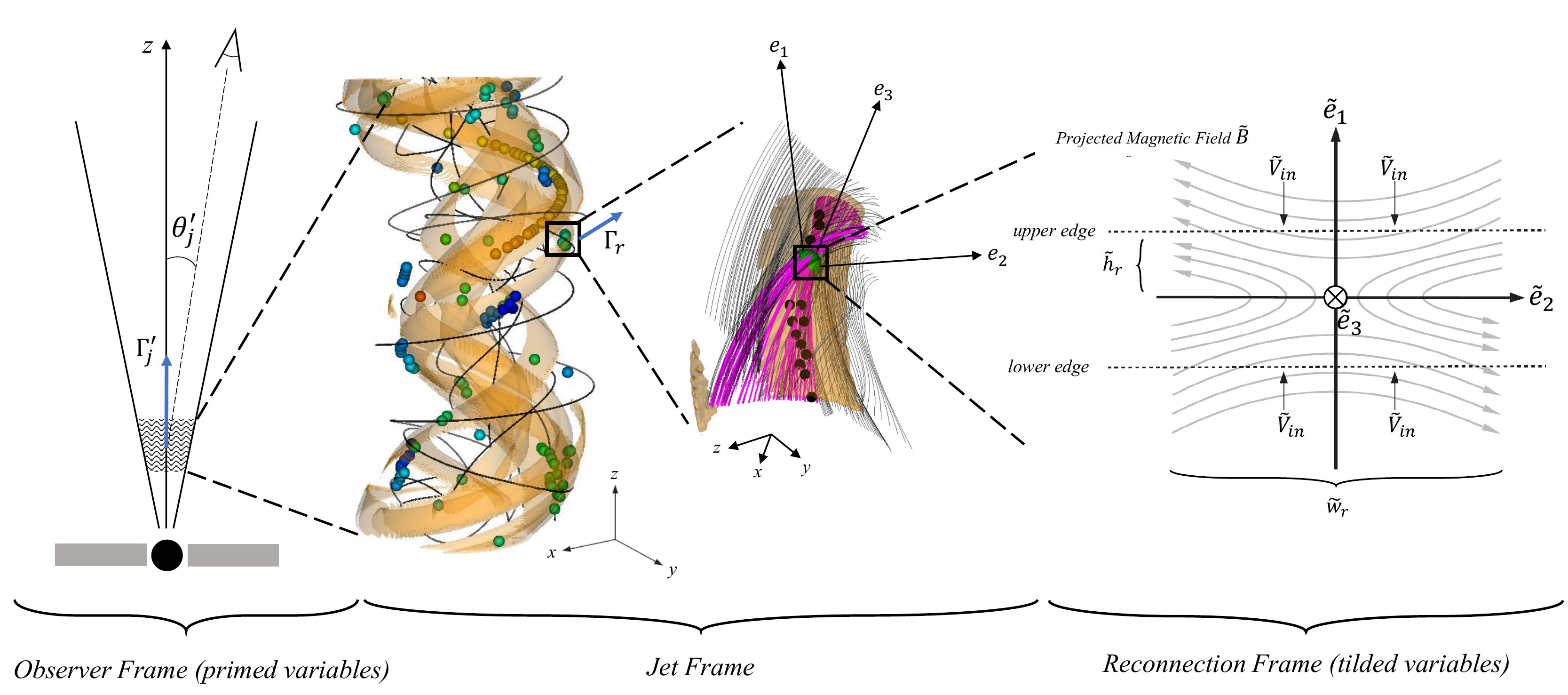}
	\caption{ The first sketch (from the left to the right) shows the relativistic jet at the observer reference frame (primed variables). The hatched region corresponds to the simulation domain moving with an angle $\theta_j^{\prime}$ to the line of sight and with a bulk Lorentz factor $\Gamma_j^{\prime}$. In physical units, this region, identified as the dissipation region of the jet, has physical dimensions of the order of 0.1 pc. The second and third figures show the evolved structures of the simulation at the jet reference frame. The colored circles correspond to reconnection sites and the streamlines correspond to the magnetic field. Finally, the last sketch 
depicts the details of a reconnection region in a coordinate system at the reconnection reference frame (tilded variables), as used in the reconnection search algorithm. From \cite{kadowaki_etal_2021}. }
	\label{fig:sample_lic}
  \end{center}
\end{figure}

In \cite{medina_etal_2021,kadowaki_etal_2021}, we performed 3D RMHD simulations of a mildly relativistic  blazar jet employing the RAISHIN code \cite{
singh_etal_16} 
with an initial helical magnetic field configuration and a magnetization parameter $\sigma = B^2/\gamma^2 \rho h  \sim 1$ (where $B$ is the magnetic field, $\gamma$ is the Lorentz factor, $\rho$ is the density and $h$ is the
specific enthalpy).  A precession perturbation induced the growth of the CDKI.
Figure \ref{jet_points} presents snapshots of the simulated jet. We note the development of a wiggling structure along the jet spine as the CDKI grows. This distortion drives turbulence in the plasma and   then  the magnetic field lines undergo fast reconnection at different scales, in agreement with the turbulent reconnection theory \cite{lazarian_vishiniac_99}. The colored circles characterize the reconnection sites which are mathematically identified using a reconnection search algorithm 
\cite{kadowaki_etal_18,kadowaki_etal_2021}. 
Figure \ref{fig:sample_lic} schematically shows how the algorithm identifies these reconnection regions. We find fast reconnection events with average reconnection velocities $V_{rec}/V_A \simeq  0.05$. The distributions of the reconnection events and of the magnetic field strength after the growth and saturation of the tubulence follow a turbulent spectrum \cite{kadowaki_etal_2021}. 
These reconnection regions with fast rates at several scales are a key ingredient to allow for  an efficient Fermi like particle acceleration \cite{dalpino_lazarian_2005}.

\begin{figure}
  \centering
	\includegraphics[scale=0.06]{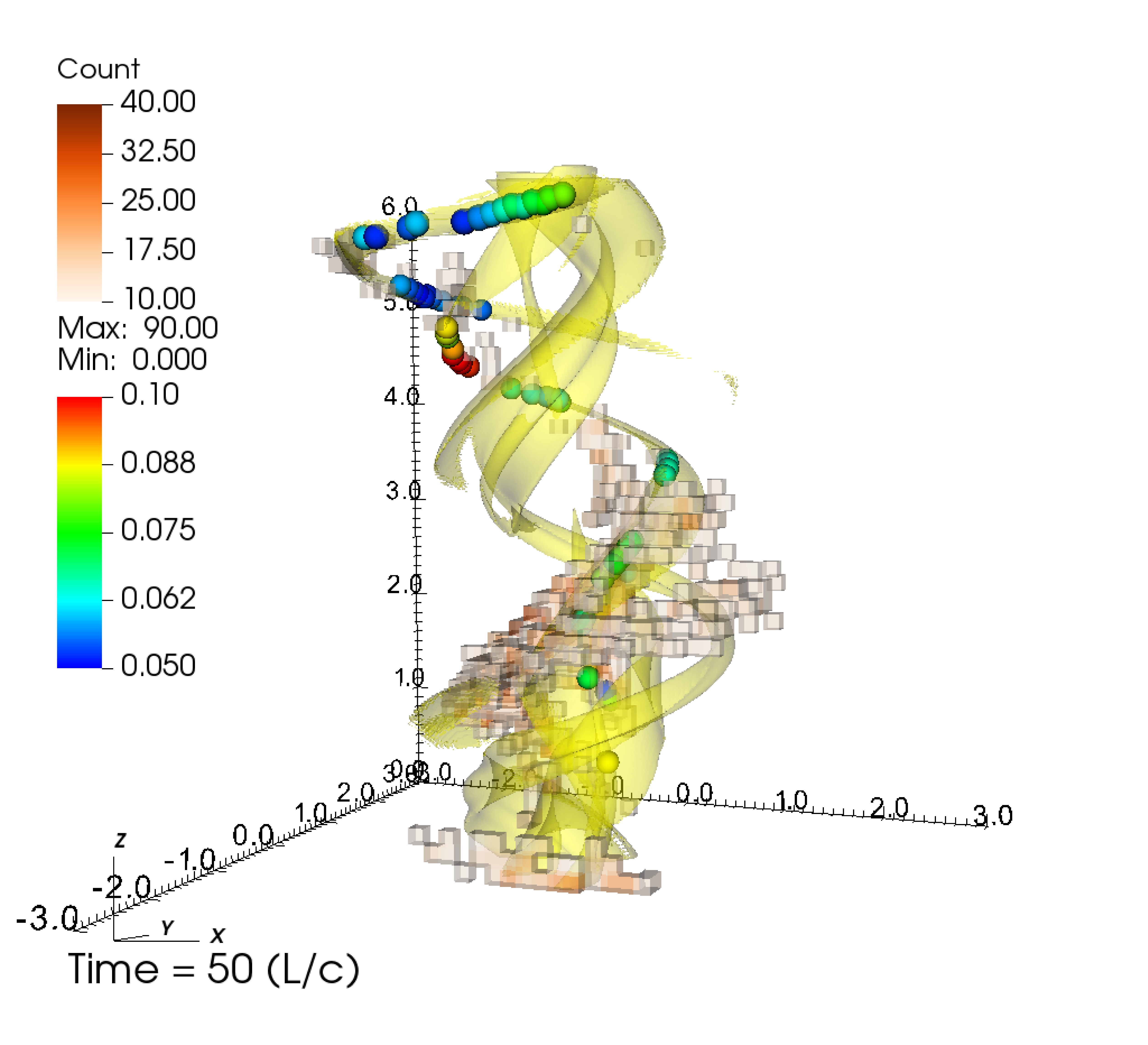}
    \includegraphics[scale=0.5]{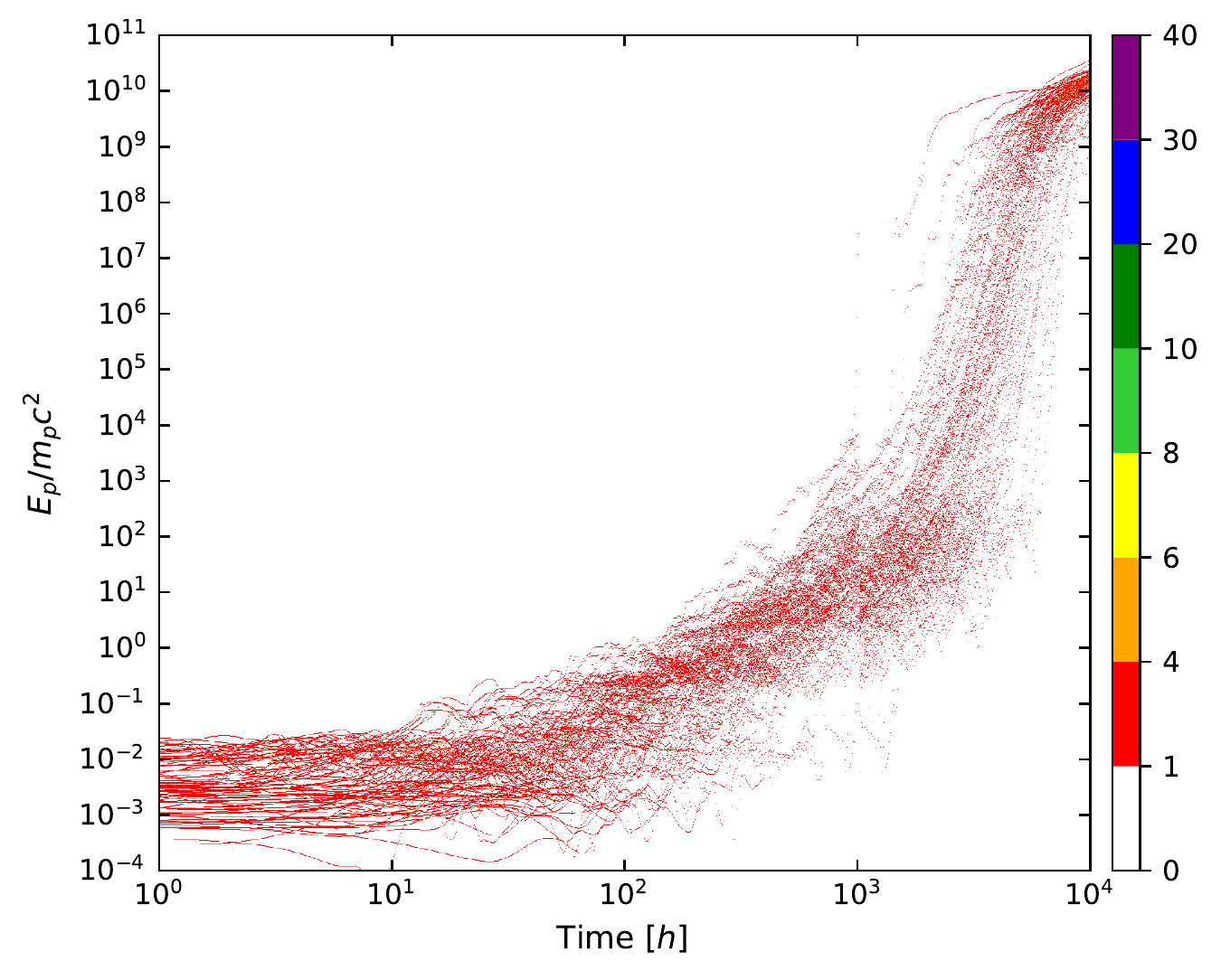}
\caption{Left panel:
3D distribution of accelerated particle positions (square symbols)  in an evolved  snapshot of the jet model after the CDKI and the turbulence achieve a nearly steady state. Only  particles accelerated with energy   $E_p > 10^{- 1} m_p c^2 $ in  the exponential regime (see right panel) are depicted.
The circles correspond to the positions of fast  magnetic reconnection sites.
The isosurfaces of  the current density intensity at half maximum $|J|$ (yellow color) are also depicted ($J_{max}/2 \sim 1.25$). 
Righ panel: kinetic energy evolution, normalized by the proton rest mass energy, for
the particles injected in the steady state snapshot $t = 50 \, L/c$ of the jet with a background magnetic field at the jet axis $B\sim 10$ G. (see \cite{medina_etal_2021} for details).}
	\label{ent46}
\end{figure}

When turbulence  achieved saturation and a  nearly stationary state, we  injected in the simulated jet   hundred to thousand protons  with initial Maxwellian velocity distribution ($10^{-3} m_p c^2$, where $m_p$ is the proton mass and $c$ is the speed of light). 
The left panel of Figure \ref{ent46} shows how the accelerated particles (orange squares) correlate with the magnetic reconnection layers (characterized by high current density)  and the sites of fast reconnection (colored circles). 
 The right panel of Figure \ref{ent46}  depicts the energy evolution of the  particles in the  nearly steady state jet
 with a maximum magnetic field   $B_0 \sim 10$ G. 
 Similarly to previous studies of test particles in single currents sheets \citep{kowal_etal_2011, kowal_etal_2012, delvalle_etal_16},   the injected particles in this simulation, after an initial slow drift in the varying background field, undergo an exponential growth in their kinetic energy.  This is due to the stochastic Fermi acceleration in the reconnection sites, as described in Section 1.  
 This regime  lasts for several hundred hours until a saturation energy around  $\sim 10^9 m_p c^2$ or $\sim 10^{18}$ eV.
During this time, the particles interact with magnetic fluctuations from the small resistive scales up to the injection scales of the turbulent structures driven by the CDKI,  of the order of the distorted jet diameter.
 Beyond this value, their energy increases further up to  $\sim 10^{20}$ eV, but in a slower, linear rate regime. At the end of the exponential acceleration the particles Larmor radius  has  achieved a value comparable to the  jet diameter. Above this energy and Larmor radius, the particles are no longer confined by the magnetic discontinuities. Still, they undergo further acceleration due to magnetic drift in the large scale fields but at a linear rate. \footnote{For a background magnetic field $B_0 \sim 0.1$ G, we find that the particles are exponentially accelerated by reconnection  up to  $\sim 10^{16}$ eV and further accelerated by drift up to  $\sim 10^{18}$ eV, i.e., for a field 100 times smaller, the maximum energy growth is 100 times smaller, as expected for the same maximum Lamor radius.} 
 
 The particles energy distribution develop a power-law spectrum with spectral index $p \sim -1.2$, which is compatible with previous MHD and PIC studies.  This hard spectrum reflects the absence of the particles feedback into the jet plasma and  radiative losses which are not accounted for in such test particle simulations. Nevertheless, these losses are not expected to alter substantially the maximum energy achieved by the particles. 

 The simulations also provide  the particle acceleration rate due to reconnection. It has  a weak dependence on the particles energy $r_{acc} \propto E^{-0.1}$,  which is comparable to earlier studies \citep[e.g.,][]{kowal_etal_2011,kowal_etal_2012,delvalle_etal_16}, and expected in a Fermi process with nearly exponential energy growth in time. 
The results from Figure \ref{ent46}
imply  a total acceleration  time (including  the  exponential acceleration regime by reconnection plus the slower final drift acceleration) of a few  $\sim 1000$  hr. During this time, particles have  traveled  a total length  of the order of   $\sim  10^{-1}$ pc along the jet. If we consider only the time elapsed during the  acceleration exponential regime, the length scales are even smaller ($\sim  10^{-2}$ pc). These  physical length scales
characterize the size of  the turbulent induced reconnection dissipation  region in the jet where particles are accelerated and, within these (time and length)  scales, the physical conditions in a real system are not expected to change substantially, except for the dissipation of the magnetic energy. It is interesting  to note that these scales are compatible with  estimates of the size of the reconnection dissipation/emission region inferred in  other blazar  studies. 

\begin{figure}
  \centering
  \includegraphics[scale=0.3]{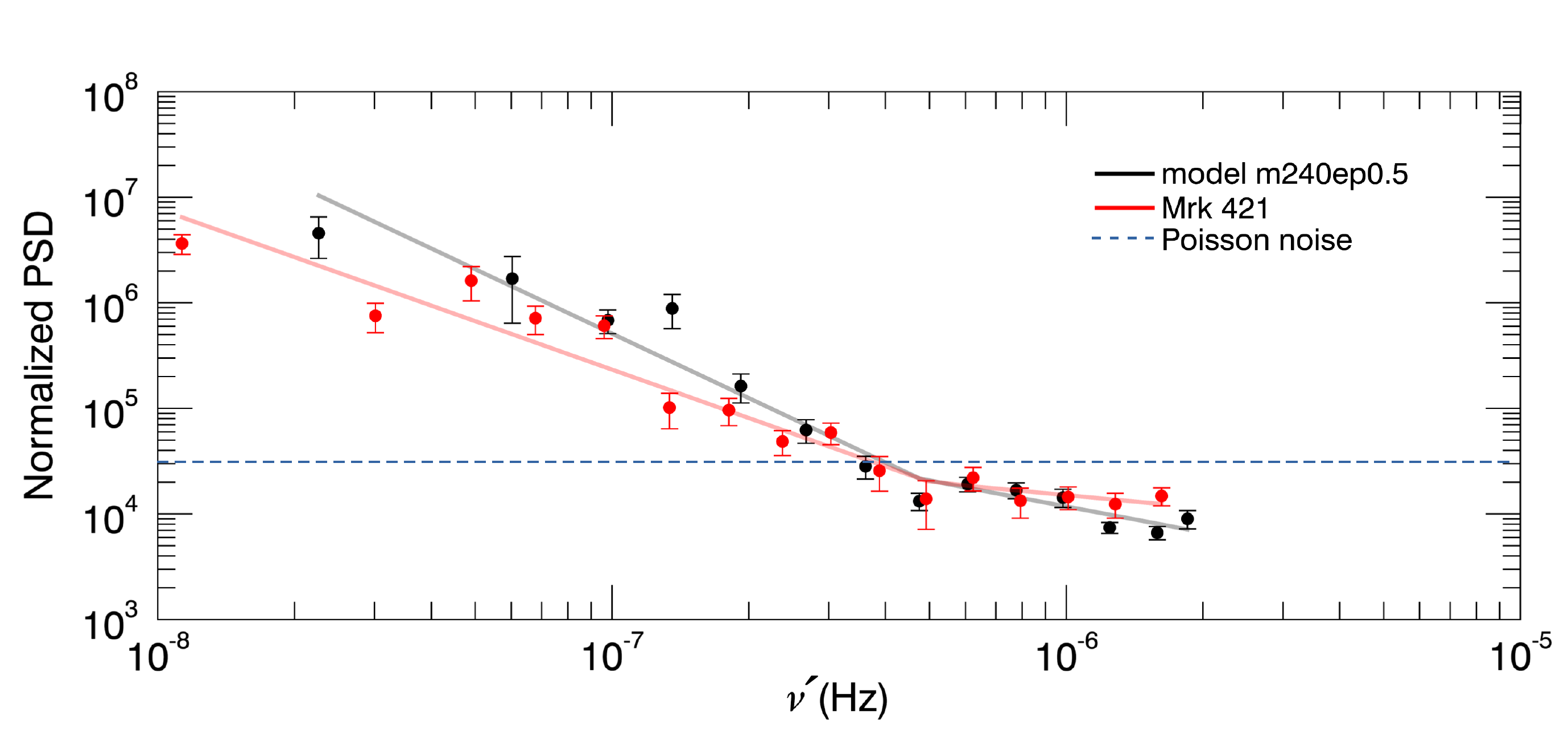}
  \caption{The diagram compares the normalized power spectral density (PSD) of the photon fluxe obtained from the light curve of MRK~421 \citep[red line, obtained from][]{kushwaha_etal_17} and that from our sumulated  jet model  (black line) in the observer frame. The blue traced line corresponds to the Poisson noise level for MRK 421. From \cite{kadowaki_etal_2021}.}
  \label{fig:psd}
\end{figure}

In summary, our study has demonstrated from first principles, by means of 3D RMHD simulations with the characteristics of blazar jets, that these can produce UHECRs. The energies obtained, well beyond the expected PeVatrons, are more than sufficient to produce TeV gamma-rays and neutrinos in these systems. Our results may also explain the time flux variability in these systems. As an example, in Figure \ref{fig:psd} we compare the observed power density spectrum (PDS) obtained from  the light curve of the blazar MRK 421  \cite{kushwaha_etal_17,kadowaki_etal_2021} in the energy band  $0.1 - 300$  GeV, with our simulations.   From the integrated  magnetic reconnection power obtained from the simulated jet, we constructed the synthetic light curve and the PDS that is shown in Figure \ref{fig:psd}.  
 These results suggest that the turbulent fast magnetic reconnection driven by CDKI  is a plausible process behind the daily time-variability observed in the GeV band in this blazar.

In  more recent work,  we have also used the results described above to build a multi-zone lepton-hadronic model based on particle acceleration by magnetic reconnection, in order  to reproduce the spectral energy distribution  (SED) of blazars, from radio to gamma-rays and neutrinos. We have  accounted for all the relevant  non-thermal radiative losses, including  both leptonic  (Synchrotron and inverse Compton) and  hadronic processes  (photo-pion production out of interactions with background photon fields, photo-hadron  and photon-photon interactions with pair production). The results,  applied, in particular, to the blazar  TXS0506$+$056, indicate good agreement with the observations (see more details in Rodriguez-Ramirez et al., this Conference).

\acknowledgments
 We acknowledge support from the Brazilian Funding Agencies FAPESP (grant 13/10559-5) and CNPq (grant 308643/2017-8).

\bibliographystyle{JHEP}
\bibliography{bibliography.bib}



%
%
%

\end{document}